\pdfoutput=1
\documentclass[twocolumn,aps,prb,10pt,floatfix,superscriptaddress]{revtex4-2}
\newcommand{\up}{\uparrow}
\newcommand{\down}{\downarrow}

\usepackage{graphicx}
\usepackage{bm}
\usepackage{dcolumn}
\usepackage{amssymb}
\usepackage{amsmath}
\usepackage{amsbsy}
\usepackage{amsfonts}
\usepackage{epsfig}
\usepackage{graphicx}
\usepackage{multirow}
\usepackage{stackrel}
\usepackage{paralist}
\usepackage[rgb]{xcolor}
\usepackage[textsize=tiny,textwidth = 1.6 cm]{todonotes}

\usepackage[percent]{overpic}
\usepackage{hyperref}
\usepackage{tikz}
\usetikzlibrary{intersections} 
\usetikzlibrary{positioning,calc}
\usepackage[caption=false]{subfig}
\usepackage[normalem]{ulem}

\tikzset{
    right angle quadrant/.code={
        \pgfmathsetmacro\quadranta{{1,1,-1,-1}[#1-1]}     % Arrays for selecting quadrant
        \pgfmathsetmacro\quadrantb{{1,-1,-1,1}[#1-1]}},
    right angle quadrant=1, % Make sure it is set, even if not called explicitly
    right angle length/.code={\def\rightanglelength{#1}},   % Length of symbol
    right angle length= 1 ex, % Make sure it is set...
    right angle symbol/.style n args={3}{
        insert path={
            let \p0 = ($(#1)!(#3)!(#2)$) in     % Intersection
                let \p1 =  ($(\p0)!\quadranta*\rightanglelength!(#3)$),
                \p2 = ($(\p0)!\quadranta*\rightanglelength!(#2)$) in %$(\p0)+\rightanglelength/{veclen((#2)-(\p0)))}*((#2)-(\p0))$ in % Point on perpendicular line
                let \p3 = ($(\p1)+(\p2)-(\p0)$) in  % Corner point of symbol
            (\p1) -- (\p3) -- (\p2)
        }
    }
}

\newcommand{\cancel}[1]{}%{\sout{#1}} %{}%
\DeclareMathOperator{\Tr}{Tr}

\DeclareMathOperator{\sgn}{sgn}
\definecolor{NewColor}{rgb}{1,0,0}
\definecolor{myRed}{rgb}{1,0,0}
\definecolor{myGreen}{rgb}{0.2,0.6,0.2}
\definecolor{myBlue}{rgb}{0,0,1}

\graphicspath{{../Graphics/}{../Graphics/Publication/}}

\newcommand{\CO}[1]{\textcolor{red}{}}

\newlength{\noLengthl}

\renewcommand{\vec}[1]{{\boldsymbol{#1}}}

\DeclareMathOperator{\Pf}{Pf}

\begin{document}
% -------- Title -----------
\title{
Complex magnetic ground states and topological electronic phases of atomic spin chains on superconductors
}

\author{Jannis Neuhaus-Steinmetz}
\affiliation{Department of Physics, University of Hamburg, 20355 Hamburg, Germany}
\author{Elena Y. Vedmedenko}
\affiliation{Department of Physics, University of Hamburg, 20355 Hamburg, Germany}
\author{Thore Posske}
\affiliation{I. Institute for Theoretical Physics, University of Hamburg, D-20355 Hamburg, Germany, The Hamburg Centre for Ultrafast Imaging, Luruper Chaussee 149, 22761 Hamburg, Germany.}
\author{Roland Wiesendanger}
\affiliation{Department of Physics, University of Hamburg, 20355 Hamburg, Germany}

\begin{abstract}
Understanding the magnetic properties of atomic chains on superconductors is an essential cornerstone on the road towards controlling and constructing topological electronic matter. Yet, even in simple models, the magnetic ground states remain debated. Ferromagnetic (FM), antiferromagnetic (AFM), and spin spiral configurations have been suggested and experimentally detected, while non-coplanar phases and complex collinear phases have been additionally conjectured. 
Here, we resolve parts of the controversy by determining the magnetic ground states of chains of magnetic atoms in proximity to a superconductor with Monte-Carlo methods. We
confirm the existence of FM, AFM and spin spiral ground states,
exclude non-coplanar phases in the model and clarify the parametric region of a $\up\up\down\down$-phase. We further 
identify a number of novel complex collinear spin configurations, including the periodic spin configurations $\up\up\up\down$, and $\up \up \up \down \up \down \down \down \up \down$, which are in some cases combined with harmonic and anharmonic spirals to form the ground state. 
We topologically classify the electronic structures, investigate their stability against increasing the superconducting order parameter, and explain the complex collinear order by an effective Heisenberg model with dominant four-spin interactions.
\end{abstract}

\maketitle

\section{Introduction}
One-dimensional systems
in proximity to s-wave superconductors have recently been extensively investigated as candidates for topological superconductivity \cite{Kitaev2001UnpairedMajoranaFermionsInQuantumWires,Kitaev2009,Ryu2010}. 
This includes experiments and calculations on semiconducting nanowires in a magnetic field \cite{Zuo2012,Ronen2012,Oreg2010,Zhang2021}, self-organized atomic chains \cite{Perge2014}, and atomically constructed magnetic chains \cite{Kim2018,Schneider2020,Schneider2021a,Steiner} on superconducting substrates, e.g., Fe on Re \cite{Kim2018, Schneider2020} and Mn on Nb \cite{Schneider2021a}.
These systems are furthermore promising platforms for 
odd-frequency and triplet superconductivity \cite{Kashuba2017,Linder2019,Kuzmanovski2020}.
A central model for analyzing the electronic and magnetic properties of the above mentioned systems is the spinful one-band model with proximity-induced s-wave superconductivity, including local magnetic Zeeman fields and Rashba-spin-orbit coupling \cite{Oreg2010,Perge2013,Klinovaja2013,Vazifeh,Heimes2015,Hu,Minami}.
This model describes 1D systems that can exhibit topological superconductivity and host Majorana zero modes at its ends.
Despite its simplicity, there is an ongoing discussion about the magnetic ground state of such systems in dependence on its parameters.
Klinovaja et al. \cite{Klinovaja2013} and Vazifeh et al. \cite{Vazifeh} found by an effective spin model that the system self-organizes into a topological state in the limit of weak magnetic interactions. Hu et al. \cite{Hu} assumed harmonic spin spirals and identified the energetically most favorable ones among them. In contrast to this approach, Minami et al. \cite{Minami}  additionally found ground state spin configurations in non-superconducting systems that are not represented by harmonic spirals but either by collinear or by non-coplanar configurations. To this end, they performed Monte-Carlo simulations with an effective spin model at vanishingly small temperatures.
Furthermore, there are models including electron-electron interactions and continuum electron models to predict the magnetic phases in one-dimensional superconductors with magnetic impurities and Rashba-spin-orbit coupling, which point towards a stable, self-organized spiral magnetic phase giving rise to one-dimensional topological superconductivity \cite{Braunecker2009a,Braunecker2013,Braunecker2010,Braunecker2009}.\\
In this paper, we present Monte-Carlo calculations of the magnetic ground state of a 1D magnetic chain with proximity-induced s-wave superconductivity. We show that non-spiral non-collinear phases exist, and analyze how they are affected by superconductivity and how they affect the topological electronic phases of the system in return. In the limit of vanishing superconductivity, we identify magnetic phases of complex order and complex collinear phases in addition to previously known harmonic spirals and collinear phases.
Our calculations are first performed in a tight-binding model, where we consider the magnetization as a free parameter and do not limit it by any assumption about the magnetic ground state. Secondly, we introduce a computationally efficient method for approximately determining the magnetic ground states of large tight-binding systems, which we use to gain
understanding of the driving forces behind the complex magnetic states. To this end, we fit the parameters of a classical Heisenberg model to our tight-binding model, showing that four-spin interactions become relevant to understand the magnetic phases.\\
The paper is structured as follows. In Section II, we explain the tight-binding model and the methods used to identify its magnetic ground states. In Section III, we discuss the magnetic ground state and the resulting electronic topological phases. In Section IV, we introduce the classical Heisenberg model for approximately finding the magnetic ground state of a tight-binding model. Finally, in Section V, we summarize our findings and give an outlook for future research.

\begin{figure*}
          \subfloat{\includegraphics[width=1.0\linewidth]{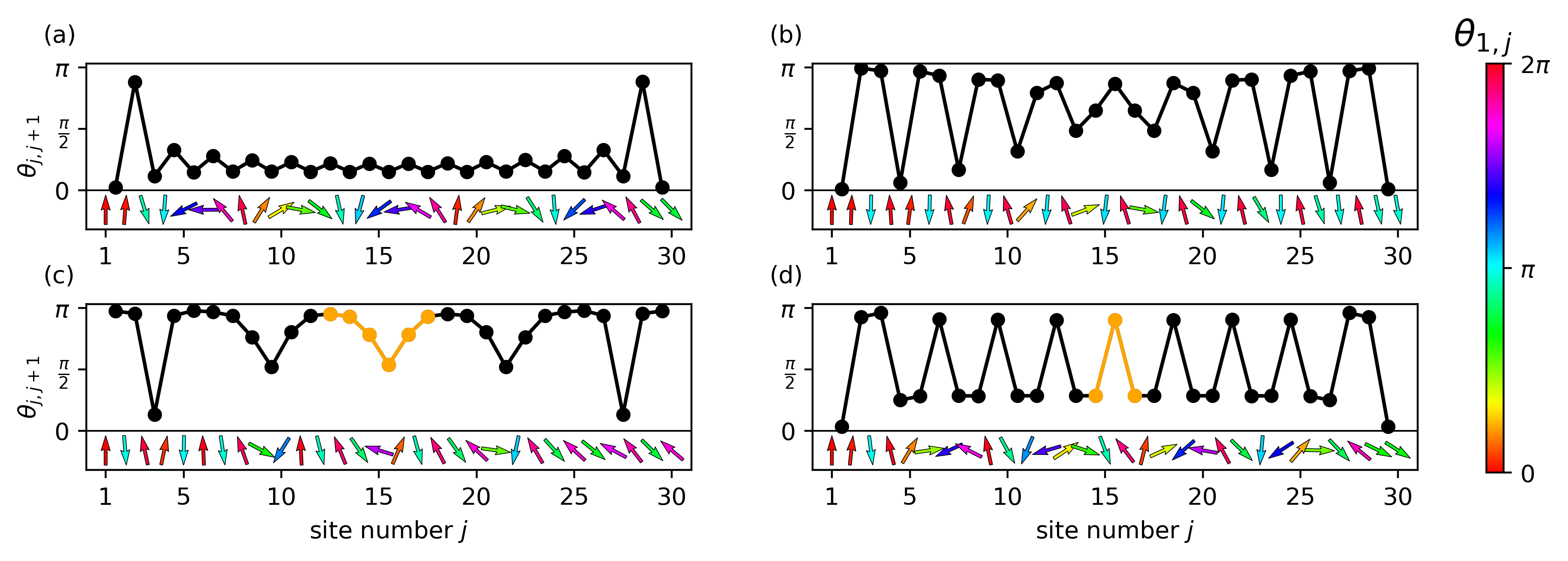}}                 
\caption{Ground state spin configurations of finite-size chains. Relative angle between neighboring spins along the chain for representative examples of ground states of finite-size chains with open boundary conditions with $\Delta=0$ and $L=30$. The periodic parts that can be used as a unit cell are marked in orange. The insets show a 2D projection of the spins. The color denotes the relative angle between the $j$-th spin and the first spin $\theta_{1,j}$. (a) $J=1.4t$, $\mu=0.5t$, (b) $J=0.2t$, $\mu=1.0t$, (c) $J=0.6t$, $\mu=1.4t$, (d) $J=1.6t$, $\mu=0.6t$.}
\label{examples}
\end{figure*}

\section{Model and Method}
We investigate a one-dimensional atomic chain with classical local magnetic moments and proximity-induced s-wave superconductivity. The system is described by the Hamiltonian
\begin{align}
H=& \sum_{j=1}^L c^{\dagger }_j\left(-J{\tau }_0\vec{m}_j\cdot \vec{\sigma }+\left(2t-\mu \right){\tau }_z{\sigma }_0+\Delta\tau_x\sigma_0\right)c_j\nonumber \\
& +\sum_{<i,j>}{c^{\dagger }_i\left(t{\tau }_z{\sigma }_0+\lambda\tau_z\sigma_y\right)c_j},\label{H_TB}  
\end{align}
with the Nambu spinor 
$c=(c_{\up },c_{\down },c^{\dagger }_{\down },{-c}^{\dagger }_{\up })$ \cite{nambu}, the coupling $J$ between a magnetic moment on a given site and the spin of an electron, the orientation of the local magnetic moments on the $j$-th site $\vec{m}_j$, the chemical potential $\mu$, the hopping amplitude $t$, the superconducting order parameter $\mathrm{\Delta }$, and the strength of Rashba-spin-orbit coupling $\lambda$. The Pauli matrices $\sigma$ and $\tau $ operate in spin and particle-hole-space, respectively. L is the length of the chain. This Hamiltonian effectively includes spin interactions mediated by the itinerant electrons and neglects direct interactions between the spins.
We choose a Rashba-spin-orbit coupling in $\sigma_y$-direction without loss of generality.
By a standard local gauge transformation $c=e^{ij\alpha\sigma_y}c'$ \cite{Braunecker2010}, the Rashba-spin-orbit coupling can be rotated into the magnetic moments $m_j'=R(2j\alpha)m_j$, where $R$ is the rotation matrix around the y-axis by an angle of $2j\alpha$. To fully rotate Rashba-spin-orbit coupling of strength $\lambda$ into the local magnetic moments, one has to set $\alpha=\arctan(\lambda/t)$, which rescales the hopping to $t'=t\sqrt{1+\frac{\lambda^2}{t^2}}$ and rotates the magnetic moments around the y-axis by an angle of $2j\arctan(\lambda/t)$. In the following, we therefore restrict our analysis to $\lambda=0$ and $t=1$. The results for non-vanishing Rashba-spin-orbit coupling can be obtained from the presented results by a backrotation of the magnetic moments by $-2j\arctan(\lambda/t)$ and rescaling of all energies by $\sqrt{1+\frac{\lambda^2}{t^2}}^{-1}$.\\
This model can host topological electronic phases despite being an s-wave superconductor depending on the magnetic configuration because the combination of hopping, s-wave pairing, and local magnetic moments can lead to an effective p-wave pairing \cite{Perge2013}. Here, we consider the magnetization as a free parameter, not limited by any a priori assumption about the magnetic ground state, and identify the energetically most favorable configuration of the magnetization $\textbf{m}_i$ for a given $J$ and $\mu$ at zero temperature with a Metropolis Monte-Carlo algorithm \cite{MC2, MC1},
and subsequently calculate ground state properties, e.g., the topological number of the electronic system and the electronic gap. 
To prevent magnetic frustration induced by incommensurate magnetic structures, we use open boundary conditions.
Details on our method are explained in Appendix A and B. The tight-binding calculations have been performed using the \textit{Kwant} code \cite{kwant}.

\section{Magnetic ground states and topological phases}
In this section, we investigate the magnetic ground state, starting with vanishing superconductivity ($\Delta=0$) and further proceeding to the superconducting case. At the end, we investigate how the magnetic states affect the electronic topological phases.\\
We start with Monte-Carlo simulations of finite chains to generate trial configurations for infinitely long chains.
Scanning through the parameter space $(J,\mu)$ with vanishing superconductivity $\Delta=0$, we find complex collinear structures at zero temperature. All ground states are globally rotationally invariant, meaning that rotating all spins simultaneously by the same angle does not affect the energy, and we do not find a spontaneous breaking of this symmetry. We observe collinear $\up\up\down\down$, $\up\up\up\down\down\down$, $\up\up\up\down$ and $\up\up\up\down\up\down\down\down\up\down$-states. Here, the short-hand notation $\up \up \down \down $ denotes that the ground state is a periodic repetition of two parallel spins and two spins that are anti-parallel to the first two spins.
We also observe structures that are dependent on finite-size effects. Representative example configurations for finite chains are shown in Fig. \ref{examples}, which shows the relative angle $\theta_{j,j+1}$ between neighboring spins along the chain. At the ends of the chains, the spins align mostly collinear and assume different structures towards the interior of the chains. Some structures (a),(b) appear to change towards harmonic spirals, but the finite-size effects can suppress this behavior (b). We also observe periodic structures, that are sequences of multiple repeating relative angles between neighboring spins (c, d). When using the inner periodic parts as unit cells for an infinite chain, the total energy of the obtained magnetic configurations is lower than that of any of the found collinear structures or spirals, implying that these structures are not caused by finite-size effects.
In contrast to Minami et al. \cite{Minami}, we did not observe any non-coplanar configurations.\\

To clarify if these configurations also exist as magnetic phases in infinite chains,
we compare the zero-temperature total energy for harmonic spirals and that of all identified collinear configurations, which we use as trial configurations for infinite chains. We do so by extracting magnetic unit cell, which are periodically expanded to do a k-space transformation. In k-space, we choose a resolution that corresponds to an effective length of 11000 atoms.

\begin{figure}
          \raisebox{-0.5 \height}{\subfloat{\includegraphics[width=0.5\textwidth]{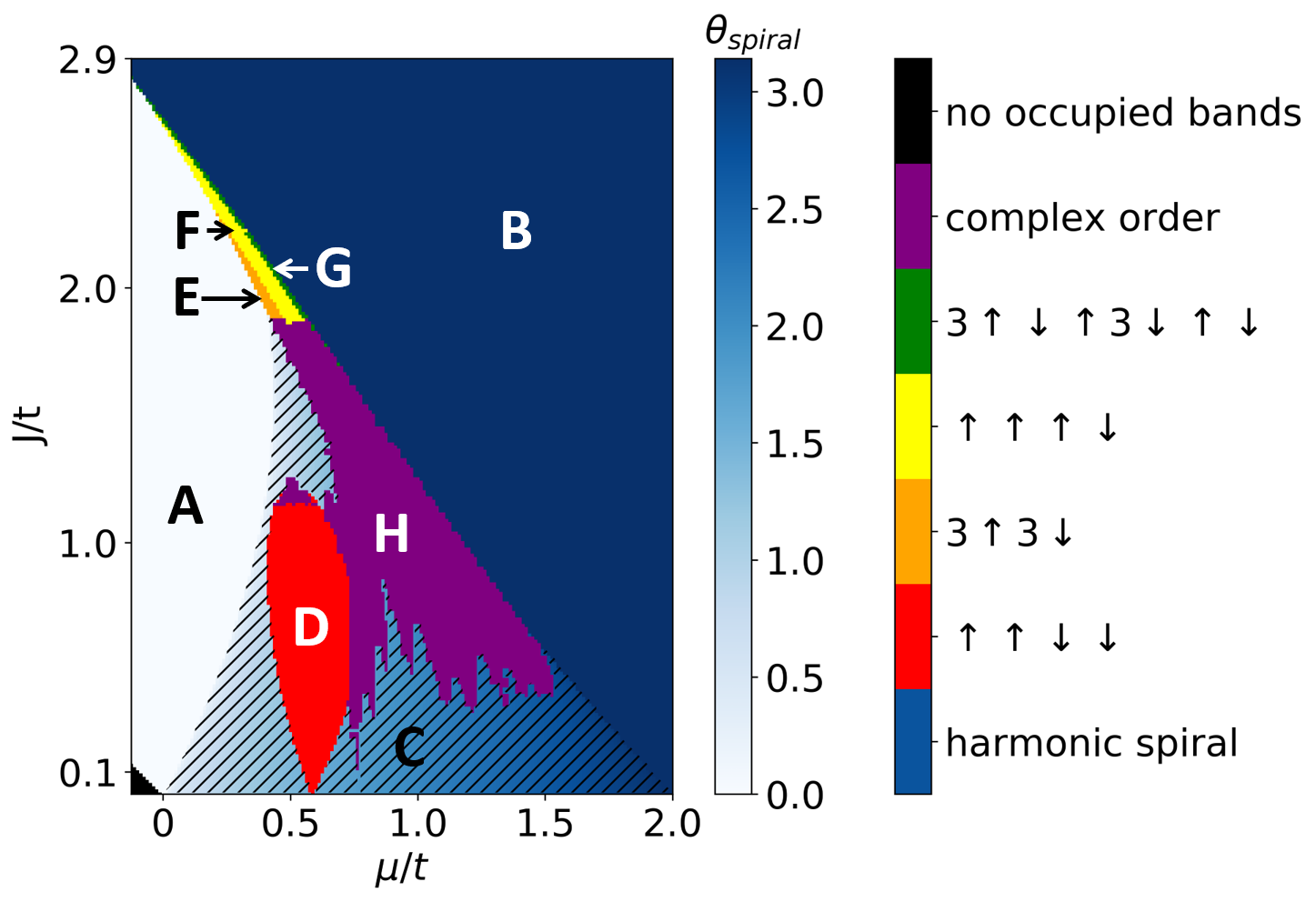}}}

\caption{Magnetic phases for vanishing superconductivity in dependence on $J$ and $\mu$ for infinite chains. The shades from white to dark blue denote a spiral, where saturation describes the spiral pitch $\theta_{\text{spiral}}$, see left color bar. The right color bar labels magnetic phases. The shortened notation 3$\uparrow$ refers to $\uparrow\uparrow\uparrow$. In the hatched area, we find a negative Majorana number $M=-1$ and an opening of a spectral gap for infinitesimal superconductivity, calculated with $\Delta=0.001t$, while the other regions remain gapless or have a positive Majorana number $M=+1$.}
\label{magn SC off}
\end{figure}

We additionally compare these configurations to results from a modified Monte-Carlo calculation that makes use of a spin basis rotation.
Fig. \ref{magn SC off} shows the magnetic phases of infinite chains identified with this method in dependence on the magnetic coupling $J$ and the chemical potential $\mu$ for vanishing superconductivity. The ground state is ferromagnetic (A) for small or negative chemical potentials $\mu$ and antiferromagnetic (B) for large $J$ or $\mu$. For $J\lesssim 0.5 t$ and $0<\mu<2t$ (C) we find a spin spiral phase, which gets interrupted around $\mu\sim0.6t$ for $J\lesssim 1.3t$ by a collinear $\up \up \down \down $-phase (D). Area E marks a collinear $\up \up \up \down \down \down $-phase.
The $\up\up\down\down$- and $\up\up\up\down\down\down$-phases have as well been reported by Minami et al. \cite{Minami}.
In addition, we identify an $\up\up\up\down$-phase (F) and an $\up\up\up\down\up\down\down\down\up\down$-phase (G) close to the $\up\up\up\down\down\down$-phase. The total energy of the collinear phases is $\sim0.05 J$ per atom lower than that of the most favorable harmonic spiral.
Similar magnetic structures have been experimentally observed with spin polarized scanning tunneling microscopy. Spin spirals (C) have been identified in Fe chains on a Re surface \cite{spirals}. Mn chains on a Nb(110) surface show ferromagnetic or anti-ferromagnetic behavior depending on their direction \cite{FM-AFM}. An $\up\up\down\down$-structure has been found in GeCu$_2$O$_4$ \cite{UUDD}.
 
Between the $\up\up\down\down$-phase and the AFM phase, we find complex non-collinear structures, that can be described as sequences of relative angles resulting in at least some non-collinear spins (H). Fig. \ref{examples} (c) and (d) represent finite-size examples of these configurations. The difference in total energy between this phase and the most favorable harmonic spiral varies from $\sim 0.03 J$ per atom close to the AFM phase to $\sim 0.002 J$ per atom close to the $\up\up\down\down$-phase.
We identified this area with a modified Monte-Carlo method, explained in Appendix D. The energetically favorable sequences of angles found with this method align well with the results from the finite-size calculations.
Finally, for $\mu <-|J|$ no bands are occupied, which is why this region is blacked out.\\

\begin{figure*}
\centering
\subfloat{\includegraphics[width=0.9\linewidth]{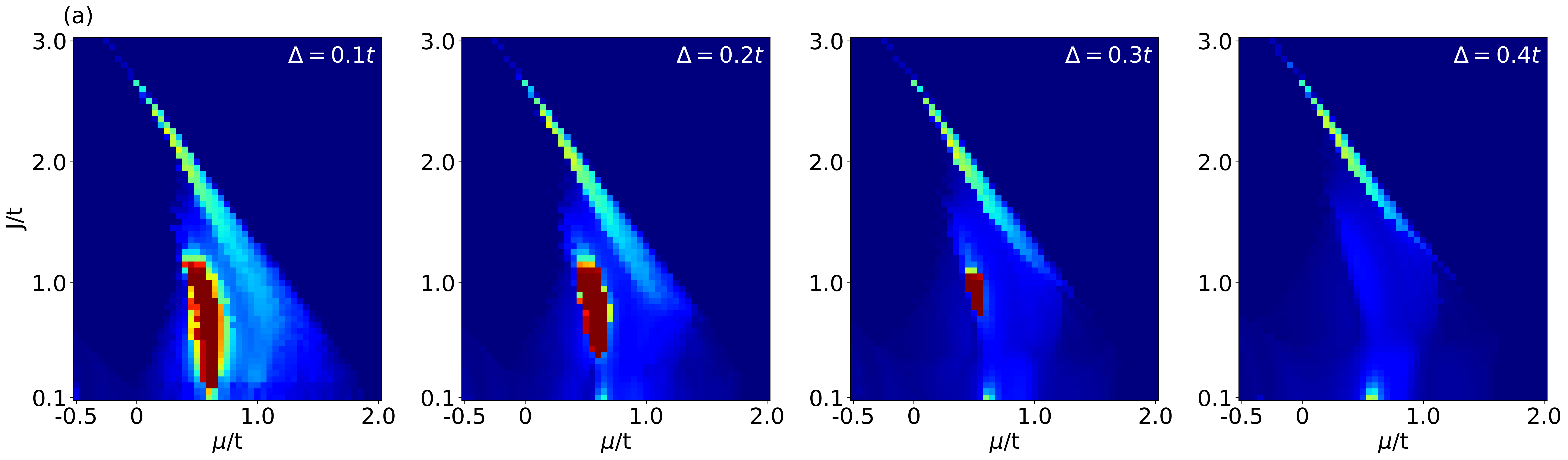}}
\subfloat{\includegraphics[width=0.054\linewidth]{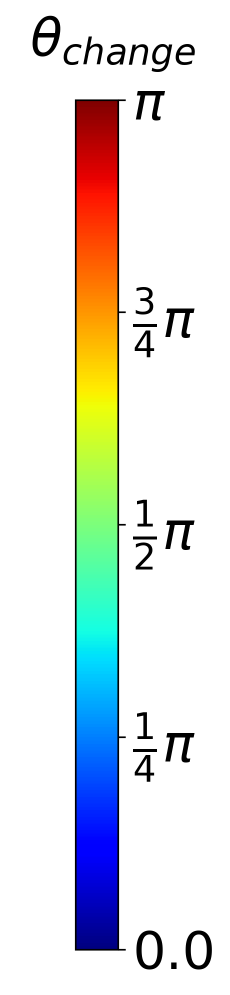}}

\subfloat{\includegraphics[width=0.9\linewidth]{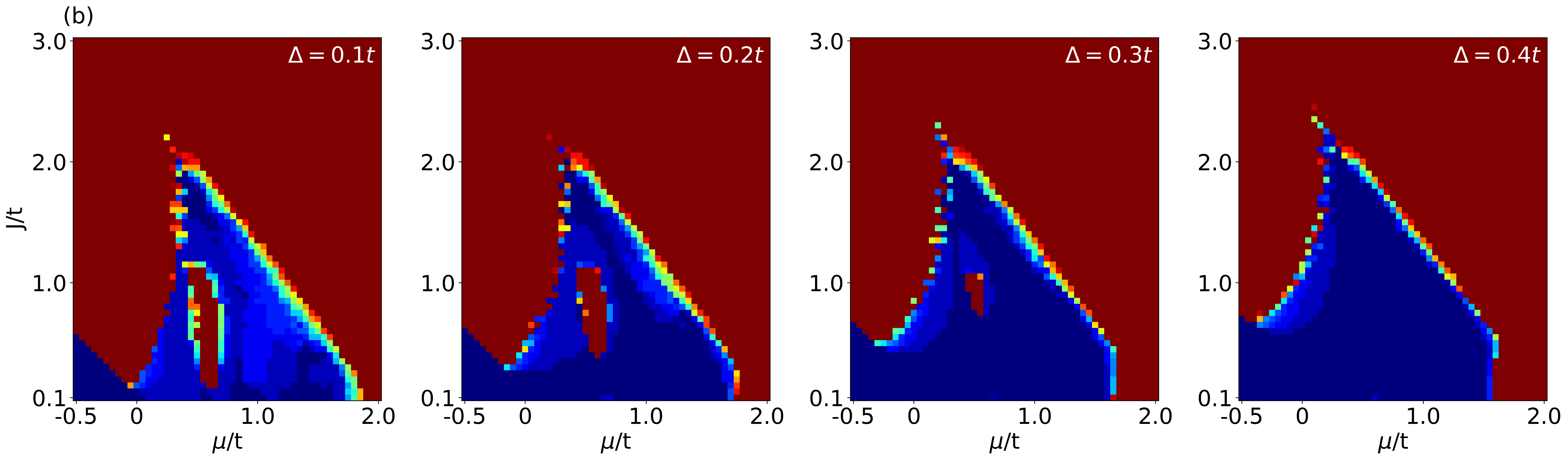}}
\subfloat{\includegraphics[width=0.054\linewidth]{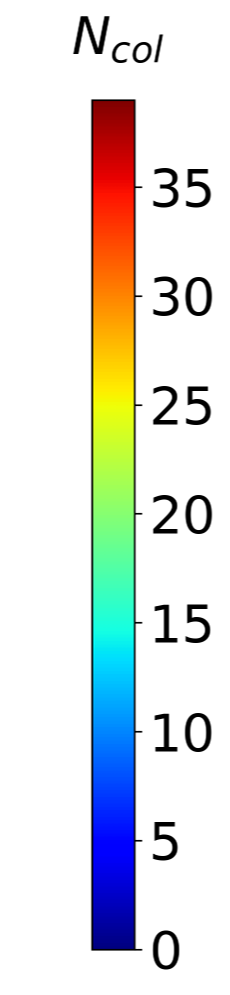}}

\subfloat{\includegraphics[width=0.9\linewidth]{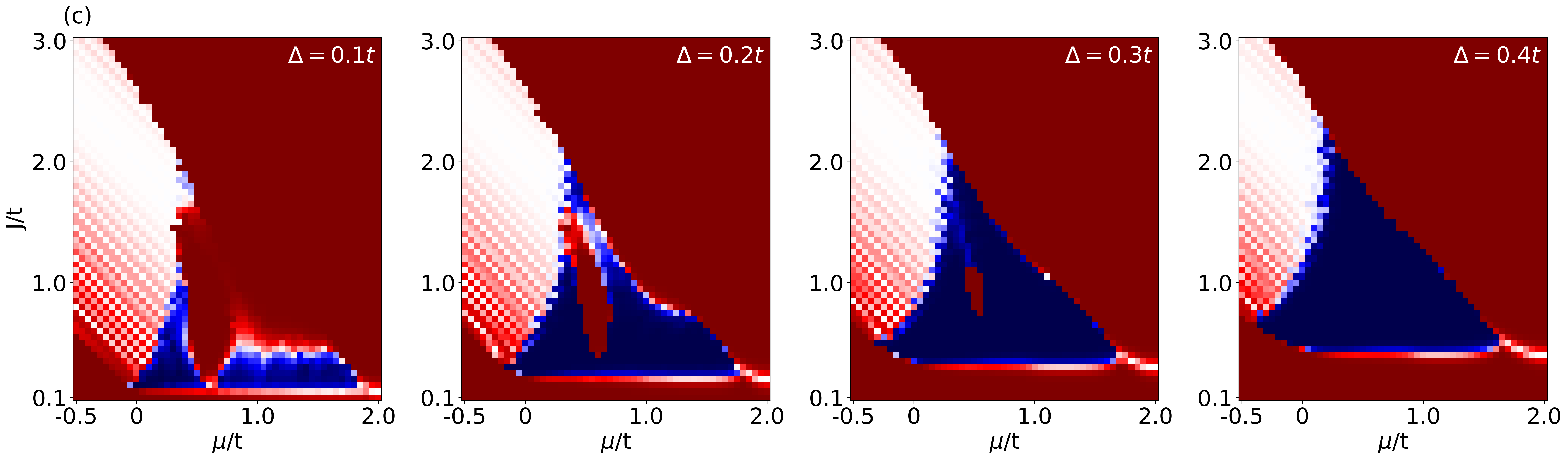}}
\subfloat{\includegraphics[width=0.054\linewidth]{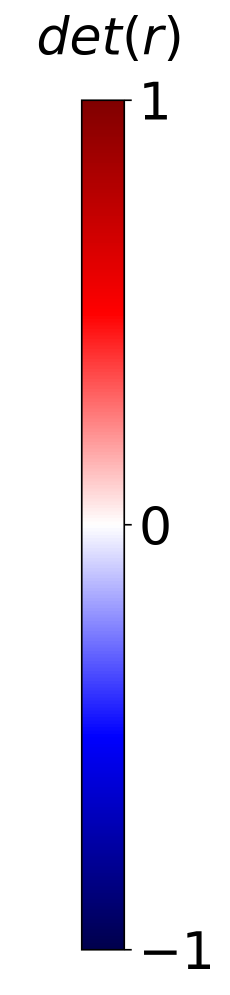}}

\caption{Magnetic properties and topological phases of finite-size chains for non-vanishing superconducting order parameters $\Delta>0$ with respect to $J$ and $\mu$, calculated for a chain of length $L=40$ with open boundary conditions. (a) $\theta_{\text{change}}$ (b) Number of collinear spins $N_{\text{col}}$. (c) The determinant of the reflection matrix $\det(r)$ in the magnetic ground state. Negative values (blue) indicate, that the system is in a non-trivial state.}

\label{finite SC}
\end{figure*}

Superconductivity, i.e, $\Delta\neq 0$, can open a spectral gap and cause the electronic system to become topologically non-trivial. Superconductivity also affects the magnetic ground states, which is investigated in the following.
Fig. \ref{finite SC} (a),(b) show the number of collinear spins $N_{\text{col}}$ and the average change of angles between neighboring spins $\theta_{\text{change}}$ for a chain of $L=40$ atoms and open boundary conditions. Here, two spins $\textbf{s}_i$ and $\textbf{s}_j$ are considered to be collinear, if $|\textbf{s}_i\cdot \textbf{s}_j|>0.99$. The quantity $\theta_{\text{change}}$ is a measure for how much a chain differs from a harmonic spiral, i.e., a chain in which the relative angle between neighboring spins remains the same along the whole chain. We define $\theta_{\text{change}}$  by
\begin{equation} \label{theta_change}
\begin{split}
\theta_{\text{change}}=\frac{1}{L-2}\sum^{L-1}_{j=1}|\arccos\left(\vec{m}_j\cdot \vec{m}_{j+1}\right)\\
-\arccos\left(\vec{m}_{j+1}\cdot \vec{m}_{j+2}\right)|.
\end{split}
\end{equation}
For small $\mathrm{\Delta }$, a significant fraction of the parameter space can not be described by harmonic spirals, i.e., the magnetic ground states have a non-vanishing $\theta_{change}$. For $\mathrm{\Delta }\gtrsim 0.35t$ the  $\up \up \down \down $-phase disappears. For $\mathrm{\Delta }\gtrsim 1.5t$, the  $\up \up \up \down \down \down$-phase and $\up \up \up \down$-phase disappear. Thus, for $\Delta\gtrsim 1.5 t$, the system converges towards harmonic spirals for all $J$ and $\mu$.

In the following, we investigate the electronic topological phases of this system.
As our model is a one-dimensional class D material with the time reversal symmetry being broken by the magnetic moments and the particle-hole symmetry squaring to one, it has a $\mathbb{Z}_2$-invariant \cite{tenfold}.
We employ the Majorana number $M$ \cite{Kitaev2001UnpairedMajoranaFermionsInQuantumWires} for the topological classification of infinitely long chains, which is
\begin{equation}
M=\sgn(\Pf(\tilde{H}(k=0)))\cdot \sgn(\Pf(\tilde{H}(k=\pi))),
\end{equation}
with the Pfaffian $\Pf$ and the k-space Hamiltonian  $\tilde{H}$ in a Majorana basis. The Hamiltonian is brought into a Majorana basis with the unitary transformation $\tilde{H}=U^\dagger H U$ with
\begin{equation}
U=\frac{1}{\sqrt{2}}\begin{pmatrix}
1 & 0 & 0 & i\\
0 & 1 & i & 0\\
0 & 1 & -i & 0\\
-1 & 0 & 0 & i
\end{pmatrix}.
\end{equation}
The system is topologically non-trivial, when it has a non-zero spectral gap and $M=-1$. 
We calculate the Majorana number $M$ for the magnetic phases shown in Fig. \ref{magn SC off} for $\Delta=0.001t$ to investigate which regions of the parameter space are non-trivial for infinitesimal superconductivity, which we expect to leave the magnetic phases unchanged.

The different magnetic ground states result in different topological electronic phases. Hu et al. \cite{Hu} found a large topologically non-trivial regime by assuming harmonic spiral ground states. Yet, the collinear and complex order phases, that we find, affect the electronic states differently than harmonic spirals and cause the electronic system to become topologically trivial, despite having an open gap. Thus, significant portions of the parameter space are in fact
topologically trivial in the ground state.
As the $\up \up \down \down $-phase (D) and the phase with complex orders (H) lie inside the harmonic spiral phase, this adds a topological phase transition that does not exist without a magnetic phase transition.\\

We further investigate the topological number of finite-size chains for larger superconducting order parameters $\Delta$. Majorana zero-modes are localized at the ends of the chain, where we find  significant finite size effects in the form of a different magnetic configuration at the boundaries than in the center of the chain. Thus, the magnetic finite-size effects can potentially affect the formation of Majorana modes even for parameters ($J, \mu, \Delta$) that would lead to non-trivial states in infinite chains. 
To calculate the topological number of finite-size chains, we use the reflection matrix $r$, following Ref. \cite{scatterD}, which is defined as the matrix, that connects an incoming mode at zero energy from an infinite lead with the reflected outgoing mode. The lead is defined by the Hamiltonian
\begin{equation}
H_{\text{lead}}=\sum_{<i,j>}{c^{\dagger }_i\left(t{\tau }_z{\sigma }_0\right)c_j}.
\end{equation}
The topological number is then calculated as 
\begin{equation}
Q=\sgn(\det(r))
\end{equation}
in a Majorana basis by using the same unitary transformation as above. This approach is equivalent to the Majorana number as defined by Kitaev for translationally invariant systems, but can be applied to systems that can not be extended to infinity \cite{scatterD}, which is the case here because of the aperiodic relaxation of the magnetization towards the ends of the chain, see Fig. \ref{examples}.\\
Fig. \ref{finite SC} (c) shows the topological phases for representative values of $\mathrm{\Delta }$ for finite chains with length $L=40$. The determinant $\det(r)$ is shown instead of $Q=\sgn\det(r)$, because $Q$ can change sign for determinants close to zero because of the numerical precision of calculating the scattering matrix. For $\Delta=0.1t$, we find topologically trivial regions, that would be expected to be non-trivial for infinite chains.
The portion of collinear aligned neighbors appears to be strongly correlated to the topological number $Q$. The more collinear neighbors there are, the less likely the system will be topologically non-trivial. As the collinear neighbors are usually found at the ends of the chain, where Majorana modes would localize, we interpret this as the magnetic finite-size effects disrupting the formation of Majorana modes. Note that absence of collinear neighbors does not guarantee non-trivial phases.

\section{Classical Heisenberg fit}

\begin{figure}       

\centering
\raisebox{-0.5 \height}{\subfloat{ \includegraphics[width=0.5\textwidth]{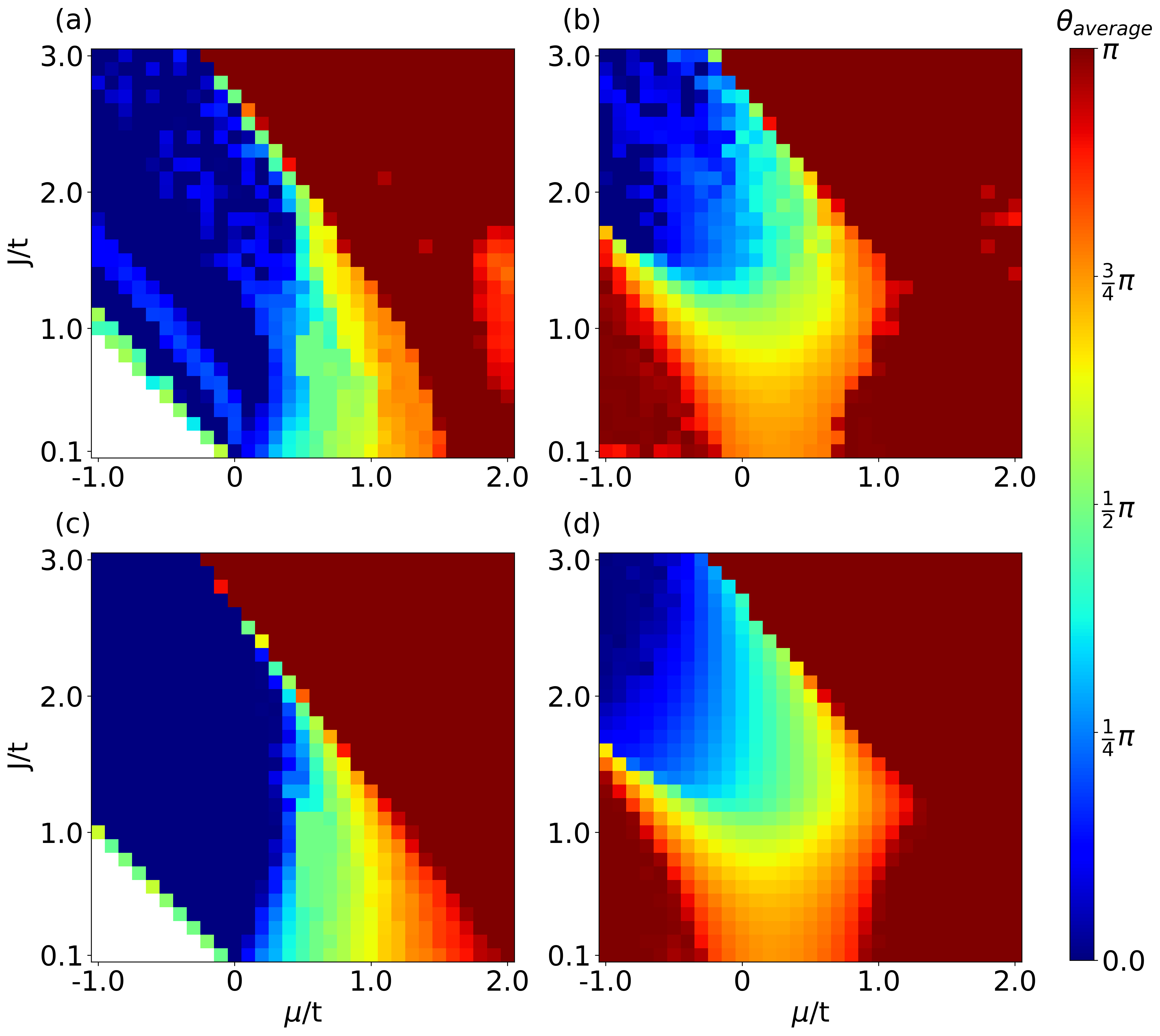}}}

\caption{Average angle between neighboring spins with respect to $J$ and $\mu$ calculated with the fitted Heisenberg model (a,b) and the tight-binding model (c,d) using Monte-Carlo. Parameters: $L=40$, $\Delta=0$ (a,c) and $\Delta=1.0t$ (b,d). Fig. \ref{variance} shows a measure of the reliability of the underlying fits.}
\label{TB vs MC}
\end{figure}

Calculating the magnetic ground states directly with a tight-binding Monte-Carlo method like the ones used in Section II and III is computationally demanding and does not grant physical understanding of the origin of the magnetic states. To improve upon both of these problems, we fit a classical Heisenberg model to the tight-binding model as follows. First, we generate a set of 3000 random configurations of the magnetization where the magnetization $\textbf{m}_i$ on each site is uniformly randomly chosen on the unit sphere. Then, we calculate the total energy of each configuration for the tight-binding Hamiltonian in Eq. \ref{H_TB} as explained in Appendix A, and construct the Heisenberg Hamiltonian $H_{\text{HB}}$ that best reproduces this data.
The Heisenberg Hamiltonian we employ is
\begin{equation}\label{H_H}
\begin{split}
H_{\text{HB}}= & \sum_{i,j}J_{i,j}\vec{s}_i\cdot \vec{s}_j+\sum_{i,j} A_{i,j} (\vec{s}_i\cdot \vec{s}_j)^2 \\
 & + \sum_{i,j,k} B_{i,j,k} (\vec{s}_i\cdot(\vec{s}_j\times \vec{s}_j))\\
& + \sum_{i,j,k,l} C_{i,j,k,l} \big([\vec{s}_i\cdot \vec{s}_j][ \vec{s}_k \cdot \vec{s}_l]+[ \vec{s}_i \cdot \vec{s}_k][ \vec{s}_j\cdot \vec{s}_l] \\
& + [\vec{s}_i\cdot \vec{s}_l][ \vec{s}_j\cdot \vec{s}_k]\big) \\
& + \sum_{i,j,k,l} D_{i,j,k,l} \big([\vec{s}_i\cdot \vec{s}_j][ \vec{s}_k \cdot \vec{s}_l] + [\vec{s}_i \cdot \vec{s}_k][ \vec{s}_j\cdot \vec{s}_l] \\
&- 2 [\vec{s}_i\cdot \vec{s}_l ][\vec{s}_j\cdot \vec{s}_k]\big).
\end{split}
\end{equation}
This Heisenberg Hamiltonian includes all isotropic 2-, 3- and 4-spin interactions \cite{tensors}. 
$J_{i,j}$,  $A_{i,j}$, $B_{i,j,k}$, $C_{i,j,k,l}$ and $D_{i,j,k,l}$ are translationally invariant, e.g., $J_{i+a,j+a}=J_{i,j}$. The summations run over all combinations of spins up to the 5th nearest neighbor. To find these parameters, we first create a sample of $N\approx 10^3-10^4$ uniformly random spin configurations and calculate the respective total energies in the tight-binding model (Eq. \ref{H_TB}). Then we fit the Heisenberg model to this sample using a least squares method.\\ 
We only include rotationally invariant terms, because the tight-binding Hamiltonian is rotationally invariant. For example, for the 2-spin interaction we include the magnetic exchange \cite{exchange} but not the Dzyaloshinskii–Moriya interaction (DMI) \cite{exchange} as DMI is not rotationally invariant. We check that the assumption of no DMI is valid, by temporarily allowing DMI in the fitting process, which consistently returns vanishing values for DMI.
We tested the quality of the Heisenberg fits by generating the magnetic ground state using the fitted Heisenberg model.
The resulting ground states are in good agreement with the results directly obtained from the tight-binding model as shown in Fig. \ref{TB vs MC}. 

\begin{figure}
	\includegraphics[width=0.5\textwidth]{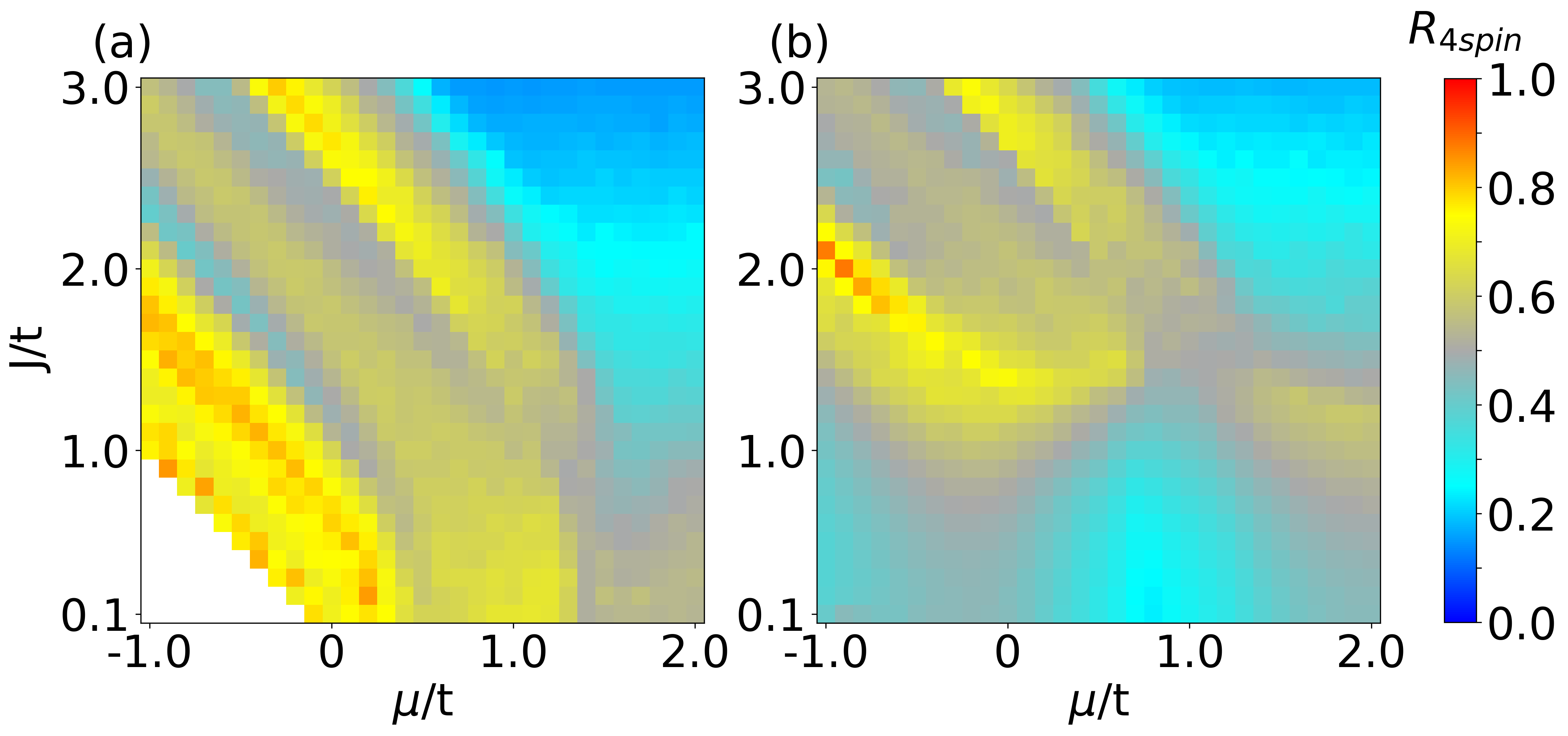}

\caption{Ratio of the 4-spin interaction $R_{\text{4spin}}$. Parameters: $L=40$, $\Delta=0$ (a), $\Delta=1.0t$ (b). Fig. \ref{variance} shows a measure of the reliability of the underlying fits.}
\label{spin4}
\end{figure}

\noindent We can understand the emergence of the complex collinear phases qualitatively with this model.
In Fig. \ref{spin4}, the ratio of the 4-spin interactions and the sum of all interactions is shown, calculated as
\begin{equation}
R_{\text{4spin}}=\frac{\sum_i (|C_i|+|D_i|)}{\sum_i(|J_i|+|A_i|+ |B_i|+|C_i|+|D_i|)},
\end{equation}
where the summation runs over all unique parameters of the Heisenberg model. This ratio shows how strong the 4-spin interaction is, relative to the whole magnetic interactions.
In the AFM phase, 2-spin interactions are dominant. In all other phases, 2-spin and 4-spin interactions are of similar strength (factor of 0.7 to 1.5). This demonstrates that higher order spin interactions are very important in this system.  Among the 4-spin interactions, the anti-symmetric terms have slightly stronger contributions than the symmetric terms by a factor of 1.2 to 2 in all non-AFM phases.
Furthermore, 3-spin interactions are vanishing in the whole parameter space. This is reasonable because the only rotationally invariant 3-spin interaction is the scalar triple product, which favors non-coplanar structures. This is in agreement with our findings from the previous section that do not indicate any non-coplanar ground states.\\
Besides granting insight into physical properties of the magnetism in superconducting atomic chains, the Heisenberg model has the advantage of being computationally more efficient as its computing time required for one energy calculation does not scale with the system size $L$ 
in the Monte-Carlo part as one only needs to consider local changes in energy. As larger systems require linearly more steps to converge in Monte-Carlo, the total computation time scales linearly with $L$ using this approach.
For comparison, calculating the total energy directly in tight-binding requires calculating all eigenvalues of an $L\times L$ matrix, which scales with $O(L^3)$ for a divide and conquer algorithm. Thus, the whole Monte-Carlo calculation in tight-binding scales with $O(L^4)$.\\
More details on this Heisenberg method can be found in Appendix C.

\section{Discussion and Conclusions}
In this study, we numerically determine the magnetic ground state of finite and infinite suspended magnetic chains with proximity-induced s-wave superconductivity, finding a number of complex collinear, complex unharmonic, and harmonic spin spiral ground states. 
For finite Rashba-spin-orbit coupling the magnetic ground states are superposed by a non-coplanar conical spiral with the y-axis as rotation axis. Here, the conical opening angle is uniformly random, reflecting the rotational symmetry of the chains for vanishing Rashba-spin-orbit coupling.
Contrary to previous results, our investigations show that harmonic spirals are not the magnetic ground state for small to medium values of the superconducting order parameter in large regions of the parameter space. Only for large superconducting order parameters $\Delta>1.5t$, the assumption of harmonic spirals as ground states holds. While the harmonic spiral phases lead to a non-trivial electronic topological phase, the other magnetic ground states result in trivial electronic topological phases.
We present an approximative method to find the magnetic ground state of tight-binding models, which scales better with system size than tight-binding calculations and grants physical insights into the magnetic interactions, by setting up a classical Heisenberg model that reconstructs the system's energy from random spin configurations. We find that the 4-spin interactions play an important role for the formation of the complex collinear phases.\\
The demonstration that simple tight-binding models host complex magnetic structures motivates further research on magnetic tight-binding models and experiments on atomic magnetic chains. Parametric regions where a small change in  parameters leads to large changes in the magnetic and electronic topological phases might be of special interest for additional research regarding the control of the location of topological boundary modes. Furthermore, our findings on magnetic ground states also facilitate experiments with spin polarized scanning tunneling microscopy as knowledge about the structure of expectable magnetic states helps in identifying magnetic states experimentally.
Finally, the presented classical Heisenberg approximation allows us to investigate the magnetic ground state of more complex and larger tight-binding models like 2D surfaces, magnetic chains on non-magnetic 3D-bulk systems, or models that account for large numbers of electronic orbitals. As long as the tight-binding model can be solved often enough to generate a sample for the fit ($N_{\text{sample}}\approx 10^3-10^4$) in reasonable computation times, it can be well approximated with the presented method.

\section*{Acknowledgments} 
J.N.-S. and R.W. gratefully acknowledge financial support from the European Union via the ERC Advanced Grant ADMIRE (project No. 786020). 

T.P. acknowledges support by the Deutsche Forschungsgemeinschaft (DFG) (project no. 420120155).

R.W. gratefully acknowledges funding by the Cluster of Excellence 'Advanced Imaging of Matter' (EXC 2056 - project ID 390715994) of the Deutsche Forschungsgemeinschaft (DFG).

\section*{Appendix A: Total energy of a superconducting system}
\renewcommand{\theequation}{A.\arabic{equation}}
\setcounter{equation}{0}
We start from a generic superconducting Hamiltonian in the BCS mean field description
\begin{equation}
H=\textbf{c}^\dagger h \textbf{c} +\textbf{c} \Delta \textbf{c} +\textbf{c}^\dagger \Delta^\dagger \textbf{c}^\dagger,
\end{equation}
where $h$ and $\Delta$ are matrices and $c=(c_1, c_2,...,c_N)^T$ is a vector containing all fermionic creation operators. This is transformed to
\begin{equation}
H=\textbf{p}^\dagger\begin{pmatrix}
h/2 & \Delta/2 \\
\Delta^\dagger/2 & -h/2
\end{pmatrix}\textbf{p} + \frac{1}{2}\Tr(h),
\end{equation}
with $\textbf{p}=(\textbf{c},\textbf{c}^\dagger)^T$ and the trace $\Tr$.\\
We then use a unitary transformation $U$ to diagonalize the Hamiltonian
\begin{equation}
\begin{split}
H=(U\textbf{p})^\dagger U 
\begin{pmatrix}
h/2 & \Delta/2 \\
\Delta^\dagger/2 & -h/2
\end{pmatrix}
U^\dagger (U\textbf{p})+\frac{1}{2}\Tr(h) \\
= \textbf{d}^\dagger \begin{pmatrix}
\epsilon/2 & 0 \\
0 & -\epsilon/2
\end{pmatrix}\textbf{d} +\frac{1}{2}\Tr(h),
\end{split}
\end{equation}
where $\epsilon$ is a matrix that contains all positive eigenvalues and $\textbf{d}=(\textbf{b}, \textbf{b}^\dagger)^T$ with the Bogoliubons $\textbf{b}$. Using fermionic algebra, the Hamiltonian becomes
\begin{equation}
H=\textbf{b}^\dagger\epsilon \textbf{b} +\frac{1}{2}(\Tr(h)-\Tr(\epsilon)).
\end{equation}
In this representation all Bogoliubons have positive energies and the ground state energy is
\begin{equation}
E_{\text{total}}=\sum_i \frac{\epsilon_i-\mu}{2}.
\end{equation}

\section*{Appendix B: Monte-Carlo Method}
\renewcommand{\theequation}{B.\arabic{equation}}
\setcounter{equation}{0}

We follow the standard Metropolis Monte-Carlo algorithm \cite{MC2}.
We start with a random spin configuration, where each spin is chosen randomly uniformly from a unit sphere. Then, in each step, one randomly chosen spin is changed into another random spin from a unit sphere. The total energy calculated in tight-binding (Appendix A) of the new spin configuration $E_{\text{new}}$ is compared to the total energy of the old configuration $E_{\text{old}}$. If the new configuration has a lower total energy it is accepted. If it has a higher total energy it is accepted with a probability equal to the Boltzmann distribution $exp(\frac{E_{\text{old}}-E_{\text{new}}}{k_B T})$. The temperature is progressively reduced close to zero temperature. Then, at low temperatures and zero temperature the spins are only updated by a small change in directions, remaining on the unit sphere. We consider the simulation to be converged, when increasing the number of steps does not reduce the total energy of the final configuration systematically. For this, we made spot checks with a ten times larger number of Monte-Carlo steps. We used $100000$ Monte-Carlo steps in the first part of the cooling, then $50000$ Monte-Carlo steps in the second part of the cooling, where only small changes are allowed, and finally $10000$ Monte-Carlo steps in the final part at zero temperature. One Monte-Carlo step represents testing a number of random spin changes equal to the number of spins in the system $L$.

\section*{Appendix C: Heisenberg fits}
\renewcommand{\theequation}{C.\arabic{equation}}
\setcounter{equation}{0}

\renewcommand{\thefigure}{C.\arabic{figure}}
\setcounter{figure}{0}

\begin{figure}
          \raisebox{-0.5 \height}{\subfloat{\includegraphics[width=0.5\textwidth]{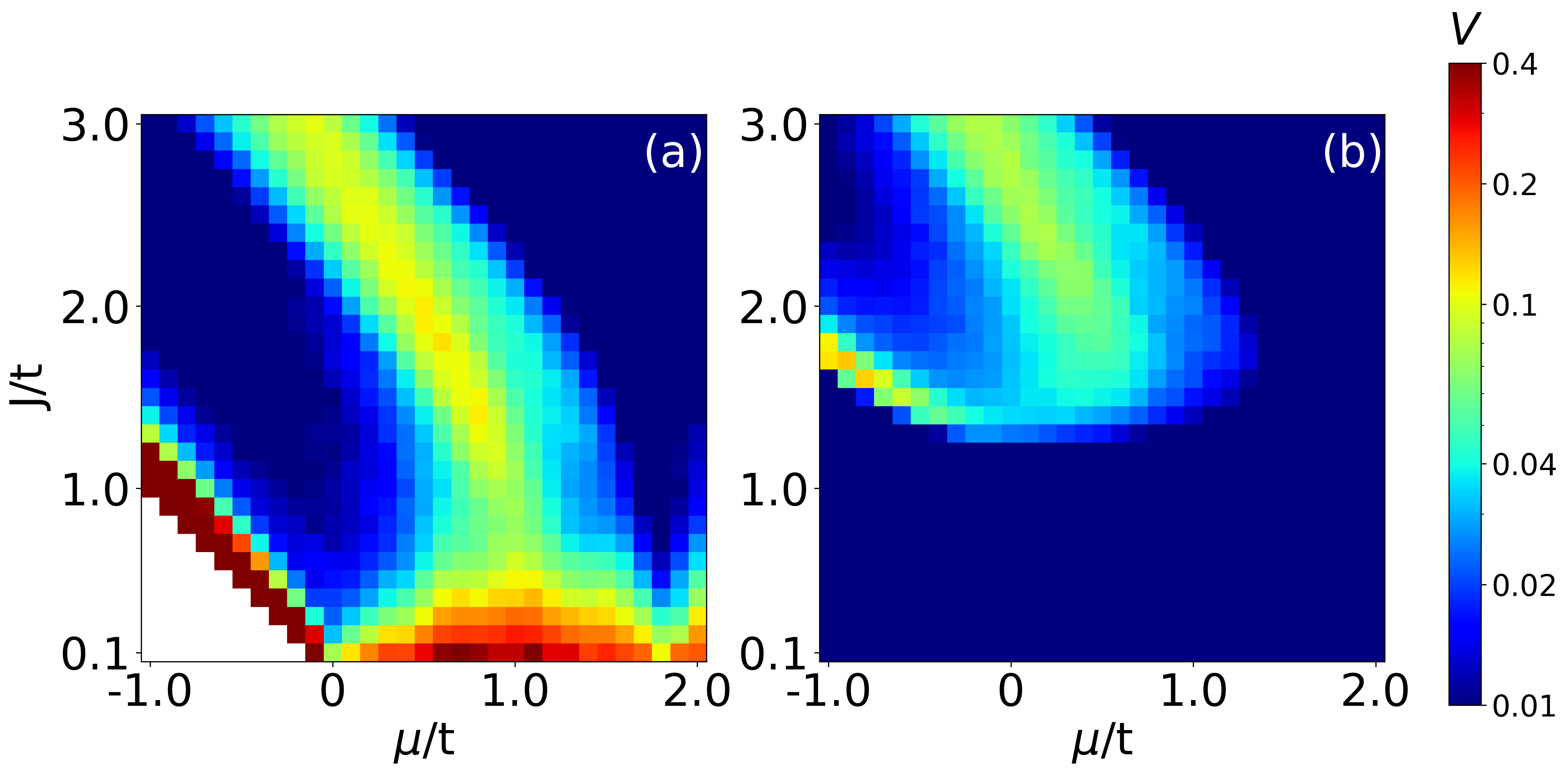}}}

\caption{Variance $V$ of the total energy difference between the Heisenberg and the tight-binding model with respect to $J$ and $\mu$ for $\Delta=0$ (a) and $\Delta=1.0t$ (b).}
\label{variance}
\end{figure}

To find the optimal coefficients $J$, $A$, $B$, $C$ and $D$ of the Heisenberg Hamiltonian (Eq. \ref{H_H} in the main text), we first create a sample of $N$ random spin configurations and calculate the respective total energies by the tight-binding model in Eq. \ref{H_TB}. We tested $N$ on the order of $10^3$ to $10^5$ and choose $N=3000$ for the calculations shown in Section IV. The required number of samples and corresponding free parameters does not scale with the system size when the Heisenberg model is chosen translationally invariant, as is the case for our model. Then we fit the Heisenberg model to this sample using the Levenberg-Marquardt algorithm, which is a least squares method.\\
To ensure that we do not overfit the results from the tight-binding calculations, we increase the sample size until the fitting parameters do not change anymore. For the calculations shown in Section IV, we set $N=3000$. Increasing the sample size to $N=50000$ results only in minimal changes to the fitting parameters. We find
\begin{equation}
\frac{var(\textbf{F}(N=3000)-\textbf{F}(N=50000)}{var(\textbf{F}(N=50000))} < 0.01
\end{equation}
for all tested $J$, $\mu$ and $\Delta$, where $\textbf{F}$ is a vector that contains all fitting parameters.\\
To judge the quality of the fits, we use a normalized variance calculated as
\begin{equation}
V=\frac{var(\textbf{E}_{\text{HB}}-\textbf{E}_{\text{tb}})}{var(\textbf{E}_{\text{tb}})},
\end{equation}
where $\textbf{E}_{\text{HB}}$ and $\textbf{E}_{\text{tb}}$ contain the energies calculated via the Heisenberg Hamiltonian and the tight-binding model for the same spin configurations, respectively. The variance for $\Delta=0$ and $\Delta=1$ is shown in Fig. \ref{variance}. For $J\approx -\mu$ the fitting process fails at $\Delta=0$, because for many spin configurations no electronic states are occupied. Within the potentially topologically non-trivial region (see Fig. \ref{finite SC}), the fitting quality is the lowest. The magnetism in that region appears to be highly complex and might require even higher order spin interactions to be fully captured by a classical Heisenberg model. Yet, regarding the general structure of the magnetic ground states, we find good agreement with the tight-binding calculations in this parameter region as well.\\

\section*{Appendix D: Monte-Carlo method for infinite chains}
\renewcommand{\theequation}{D.\arabic{equation}}
\setcounter{equation}{0}
In this section, we explain the modified Monte-Carlo method that we use to identify in which region non-harmonic spin structures with sequences of multiple different relative angles are energetically more favorable than harmonic spirals.\\
We use a spin basis rotation $R_{\theta_j}$ 
\begin{equation}
R_{\theta_j}=\begin{pmatrix}
\cos (\theta_j/2) & \sin (\theta_j/2)\\
-\sin (\theta_j/2) & \cos (\theta_j/2)
\end{pmatrix},
\end{equation}
corresponding to a rotation of the magnetization in the xy-plane.
This removes the spin directions from the onsite potential and adds a change in the spin direction to the hopping term between the $j$-th and $(j+1)$-th sites.
With this the tight-binding Hamiltonian becomes
\begin{equation}
\centering
\begin{split}
H= & \sum_j{c^{\dagger }_j\left(-J{\tau }_0\sigma_z+\left(2t-\mu \right){\tau }_z{\sigma }_0+\Delta\tau_x\sigma_0\right)c_j}\\\
&+\sum_{<i,j>}{c^{\dagger }_i\left(t{\tau }_z\otimes R_{\theta_i}\right)c_j},
\end{split}
\end{equation}
where $\theta_j$ is the relative angle between the $j$-th and the $(j+1)$th site. Writing the Hamiltonian in this way allows us to infinitely expand spin structure that can be described by a finite sequence of relative angles \cite{Martin2012}.\\
We then employ a Monte-Carlo method that varies these relative angles for unit cells of size $L=1,2,3...8$ for an infinite chain in k-space, sampling 10000 k-points. The Monte-Carlo updates are the same as described in Appendix B. We then compare the minimal total energy found for each size of the unit cell and determine the array of $\theta_i$ that achieves the lowest total energy. When two unit cells with different length coincidence in total energy (on the order of $0.001 J$), the smaller unit cell is preferred.\\
If the ground state is a harmonic spiral, FM, or AFM, one finds unit cells of size $L=1$ with this method but $L\geq 2$ for more complex structures. We also calculate how much $\theta_i$ changes along the unit cell, for unit cells with $L\geq 2$, and quantify this by the average change of the relative angle between neighboring pairs of spins:
\begin{equation}
\theta_{\text{change}}=\sum_{j=1}^{L-1} \frac{|\theta_j-\theta_{j+1}|}{L-1}.
\end{equation}
In the case of a harmonic spiral, FM or AFM one finds $\theta_{\text{change}}=0$. 
The region marked as (H) in Fig. \ref{magn SC off} reflects the parameters for which the derived unit cell has $L\geq 2$, $\theta_{\text{change}}>0.05$ and not all $|\theta_j|=\pi$, i.e., area H is a phase that can neither be described by a harmonic spiral nor by a collinear structure.

\bibliographystyle{apsrev4-2}
\bibliography{BibFile}

\end{document}